\documentclass[aps,prb,superscriptaddress, floatfix, reprint]{revtex4-2}
\usepackage{graphicx}
\usepackage{dcolumn}
\usepackage{bm}
\usepackage{amsfonts}
\usepackage{amsmath}
\usepackage{amssymb}
\usepackage{color}
\usepackage[normalem]{ulem}
\usepackage[percent]{overpic}
\usepackage{mathtools}
\usepackage{braket}
\usepackage[colorlinks=true,citecolor=blue]{hyperref}
\usepackage{tikz}
\hypersetup{colorlinks=true,citecolor=blue,linkcolor=red,urlcolor=blue}
\newcommand{\sgn}{\operatorname{sgn}}

\def\be{\begin{equation}}
\def\ee{\end{equation}}

\def\rv{{\bf r}}
\def\kv{{\bf k}}
\def\uv{{\bf u}}
\def\qv{{ \bf q}}

\begin{document}

\title{Collective excitations and screening in two-dimensional tilted nodal-line semimetals}
\author{Hamid Rahimpoor}
\affiliation{Department of Physics, Institute for Advanced Studies in Basic Sciences (IASBS), Zanjan 45137-66731, Iran}
\author{Saeed H. Abedinpour}
\email{abedinpour@iasbs.ac.ir}
\affiliation{Department of Physics, Institute for Advanced Studies in Basic Sciences (IASBS), Zanjan 45137-66731, Iran}
\date{\today}

\begin{abstract}
Topological nodal-line semimetals are characterized by symmetry-protected one-dimensional band-touching lines or loops, which give rise to their peculiar Fermi surfaces at low energies. Furthermore, if time-reversal or inversion symmetry breaking tilts the bands, anisotropic Fermi surfaces hosting electron and hole carriers simultaneously can also appear. 
We analytically investigate the linear density-density response function of a two-dimensional tilted nodal-line semimetal in the intrinsic and doped regimes. Despite the anisotropic electronic bands, the polarizability remains isotropic in our model system. We find that the plasmon dispersion in the long wavelength limit exhibits a standard behavior that is proportional to the square root of the wave vector, characteristic of two-dimensional electron liquids. Tilting tends to enhance the plasmon frequency, and the Drude weight does not depend on the carrier density at low doping levels. In these regimes, unlike the intrinsic and highly-doped ones, the static polarizability has two distinct singularities at finite wave vectors. This results in beat patterns in the Friedel oscillations.
\end{abstract}
\maketitle

\section{Introduction}\label{introduction}
Topological phases of matter are an emerging subject in modern condensed matter physics that has sparked considerable interest in the last two decades. 
Even though the starting point of these topological materials was topological insulators characterized by their gaped bulk electronic energy spectrum and symmetry-protected gapless surface states~\cite{qi2011topological,xia2009observation,hasan2010colloquium},
gapless topological materials are gaining significant attention lately.
These appealing materials are known as topological semimetals and possess zero-gap bulk states~\cite{burkov2016topological,armitage2018weyl}, and may support non-trivial surface states as well.
In topological semiemtals, the valence and conduction bands touch each other either in isolated points or along open or closed lines in the Brillouin zone (BZ). The topological band crossing in the bulk is either accidental or symmetry enforced~\cite{schnyder_julich2020}.
The low-energy excitations around the nodal points are described by the massless Dirac equation~\cite{young2012dirac,yang2014classification,burkov2011weyl,wan2011topological}, with the
representative examples being Na$_3$Bi~\cite{liu2014discovery} and Cd$_3$As$_2$~\cite{ wang2013three} for the Dirac semimetals, as well as TaAs~\cite{xu2015discovery} and TaP~\cite{weng2015weyl} for the Weyl semimetals.
Breaking either the time-reversal or inversion symmetry in Dirac semimetal results in a Weyl semimetal with two-fold degenerate point nodes in the momentum space~\cite{armitage2018weyl}. 

Dirac cones in Dirac or Weyl semimetals might be tilted~\cite{soluyanov2015type, Li_PRB2021}, and several interesting phenomena are associated with that~\cite{tan2022signatures,hosseinzadeh2023semiclassical,jalali2021tilt,faraei2020electrically}.
These tilted semimetals are normally classified according to the geometry of their Fermi surfaces at the nodal points. For instance, in so-called type-II Weyl semimetals over-tilted Dirac cones give rise to particle and hole pockets at the intersection of the Dirac cone with Fermi level~\cite{soluyanov2015type}. 

In another family of topological semimetals, namely the nodal-line semimetals (NLSM), the band crossing takes place
along a line or closed loop in the BZ~\cite{fang2016topological,fang2015topological}. A large variety of material candidates are proposed for NLSMs. 
Cu$_3$PdN~\cite{yu2015topological} and CaAgAs~\cite{wang2017topological} are among the vast number of theoretical candidates, whereas ZrSiS~\cite{schilling2017flat,topp2017surface}, NbAs$_2$~\cite{shao2019optical}, and YbMnSb$_2$~\cite{qiu2019observation} are examples of experimentally validated three dimensional (3D) NLSMs.
Similar to Dirac cones, the nodal line can be tilted too due to time-reversal or inversion symmetry breaking. Such tilted nodal lines are observed in several materials, including ZrSiS, HfSIS, and ZrSiSe~\cite{Chen_PRB2017,Shao_NatPhys2020}.

Nodal-line semimetals are investigated in two dimensions as well. Jin \emph{et al.} predicted a family of two-dimensional (2D) nodal-line semimetals MX with M=Pd, Pt; and X=S, Se, Te~\cite{jin2017prediction} using evolutionary algorithm and first-principles calculations, and Lu \emph{et al.} suggest possible realization of 2D NLSMs in a mixed honeycomb-Kagome lattice structure~\cite{lu2017two}. 

Various properties of different types of NLSMs have been extensively explored so far (see, e.g., \cite{schnyder_julich2020}, and references therein). 
Optical conductivity~\cite{barati2017optical,ahn2017electrodynamics} and thermoelectric responses~\cite{barati2020thermoelectric} of two and three-dimensional NLSMs are investigated. 
As for the collective excitations, the plasmon frequency of 3D NLSMs at long wavelength shows a $n^{1/4}$-dependence on the carrier density $n$, that is distinct from the ordinary electron liquids, and Dirac or Weyl semimetals~\cite{ yan2016collective}. 
The Friedel oscillations in 3D NLSMs exhibit an angle-dependent algebraic power-law decay~\cite{rhim2016anisotropic}. Also, a recent study~\cite{cao2022plasmons} of the collective modes of a 2D non-symmorphic NLSM with an open band-touching line, predicts a strongly anisotropic plasmon dispersion in these materials. 

In this paper, we consider an effective low-energy two-band Hamiltonian with a circular nodal loop and a linear tilt term. We investigate the zero-temperature linear density-density response function of this model system in the static and dynamic regimes at different carrier concentration levels.
 The rest of this paper is organized as follows. Sec.~\ref{sec:model} introduces the low energy effective Hamiltonian of a 2D tilted NLSM and discusses its different doping regimes. In Sec.~\ref{sec:Polarizability}, we analytically calculate the imaginary and real parts of the linear density-density response function in the intrinsic and doped regimes. The static limit behavior and screening of the charged impurity are discussed in Sec.~\ref{sec:Static screening}, while Sec.~\ref{sec:Plasmons} is devoted to the discussion of the collective density mode and the behavior of plasmon frequency at long wavelength as well as arbitrary wave vectors within the random phase approximation (RPA).  
Finally, we conclude and summarize our main findings in Sec.~\ref{sec:summary}.

\section{Model Hamiltonian}\label{sec:model}
We consider the following effective low-energy single-particle model Hamiltonian, for a two-dimensional semimetal with a circular nodal line and linear tilt  ($\hbar=1$)~\cite{ekstrom2021kerr,martin2018parity}
\begin{equation}\label{eq:hamil}
\hat{H}_{0} =  \uv \cdot \kv\ \hat{\tau}_0+ \frac{1}{2 m} (k^2 - k_0^2) \hat{\tau}_x. 
\end{equation}
Here, $\uv$ is the tilt velocity vector, $\hat{\tau}_0 $ and $\hat \tau_{x}$ are respectively the two by two identity matrix and $x$-component of the Pauli matrix, acting on the pseudo-spin (i.e., orbital) degree of freedom, $m$ is the band mass, $ k=\sqrt{k^2_x+k^2_y}$ is the magnitude of the wave vector, and $ k_0 $ is the radius of nodal loop in the absence of tilt. 
Eigenvalues corresponding to Hamiltonian \eqref{eq:hamil} are given by  
\begin{equation}\label{eq:epsilon}
\varepsilon _{\kv,s} = \uv.\kv+\dfrac{s}{2m}|{k}^2 -{k}_0^2|,
\end{equation}
where $s=+(-)$ labels the conduction (valance) band. 
Moreover, the eigenstates of our model Hamiltonian in Eq. \eqref{eq:hamil} are readily obtained as
 \begin{equation}\label{eq:eigenstates}
 \ket{\kv,s} \doteq \frac{1}{\sqrt{2}} \begin{pmatrix}
 1\\
 s\ \sgn({k}^2 -{k}_0^2)
 \end{pmatrix}.
 \end{equation}
Here, $\sgn(x) $ is the sign function.
The tilt term considered in Eq.~\eqref{eq:hamil} does not affect the eigenstates and
in the $ \uv \to 0 $ limit, we recover the model Hamiltonian for un-tilted 2D NLSMs~\cite{barati2017optical}. 
It is worth noting that in the BZ of real materials, tilted nodal lines should appear in pairs of opposite tilt velocities, either at the same point or at different valleys in the BZ. As we will see later on, our results are independent of the tilt direction, therefore any extra nodal-line will contribute to the degeneracy factor.
In the following, without losing the generality of our problem, we will take the tilt along the $x$ direction and set $ u_y=0 $. 

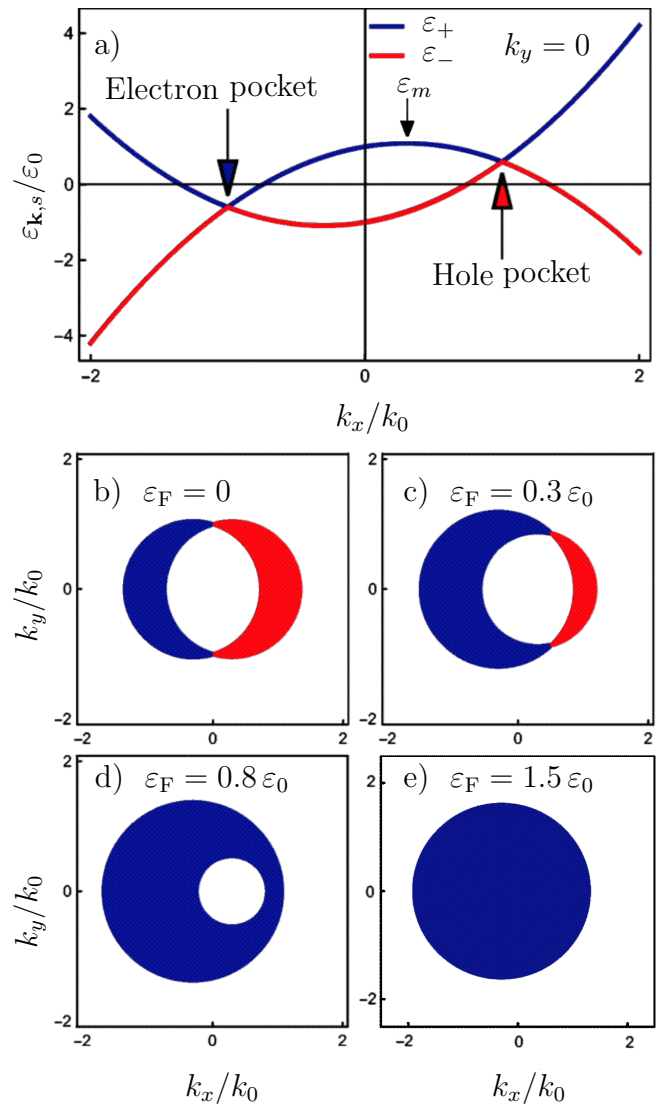
\begin{figure}
 	\centering
 	 \begin{tikzpicture}[scale=1.6]
 	\node at (0,0) {\includegraphics[width=1.0\linewidth]{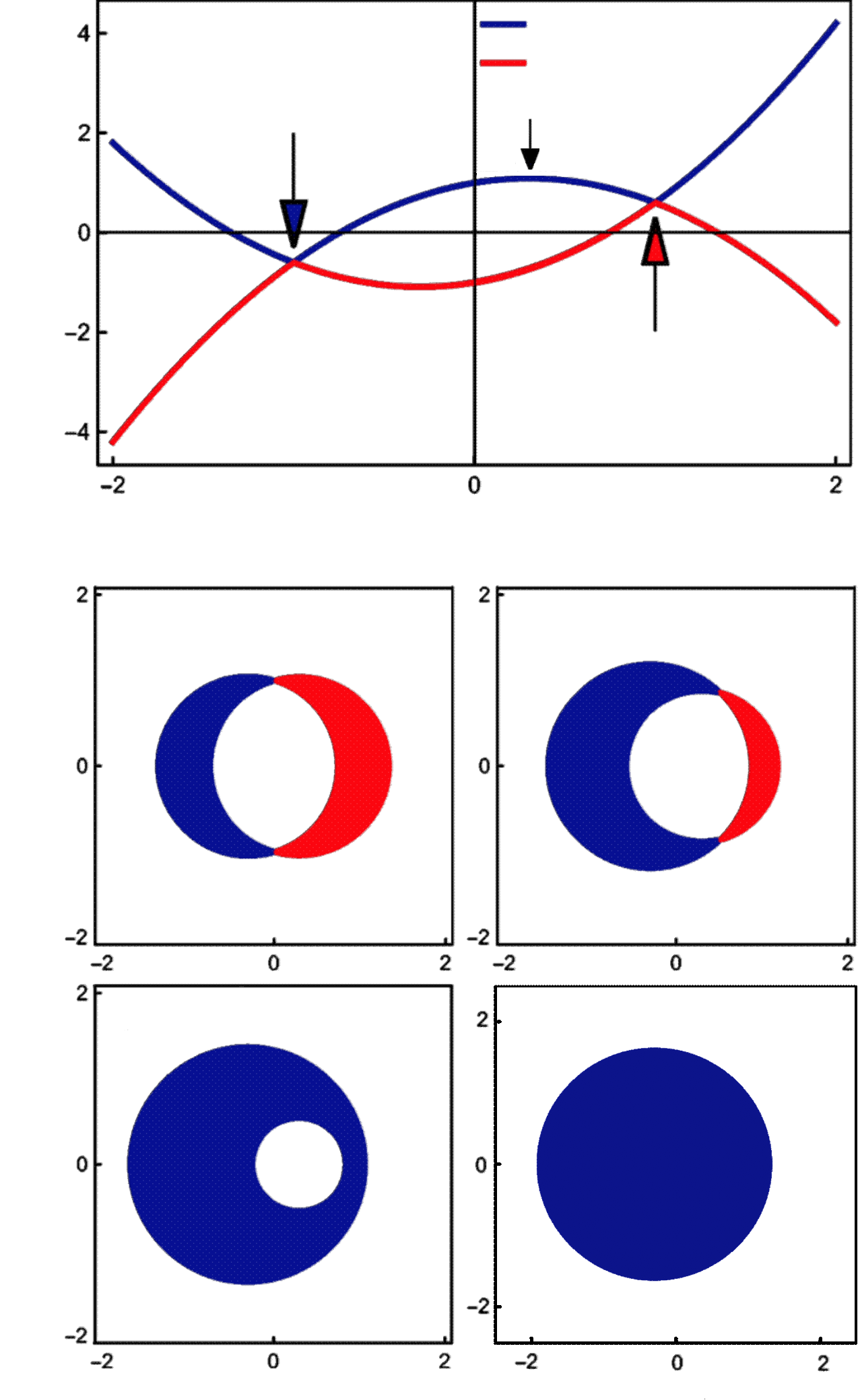}};
 	\draw (0.35,1) node [font=\large] {$ k_x/k_0 $};
 	\draw (-2.5,2.85) node [font=\large] [rotate=90] {$ \varepsilon_{\kv,s}/\varepsilon_0 $};
 	\draw (0.9,4.25) node [font=\large] {$ \varepsilon_{+} $};
 	\draw (0.9,4) node [font=\large] {$ \varepsilon_{-} $};
 	\draw (0.7,3.75) node [font=\large] {$ \varepsilon_{m} $};
 	\draw (-1.4,3.75) node [font=\large] {Electron };
 	\draw (-0.47,3.75) node [font=\large] { pocket };
 	\draw (1.1,2.2) node [font=\large] { Hole };
 	\draw (1.8,2.2) node [font=\large] { pocket };
 	\draw (1.8,4.08) node [font=\large] {$ k_y=0 $};
 	\draw (-1.85,4.1) node [font=\large] {a) };
 	\draw (-1.4,0.38) node [font=\large] { b) $~\varepsilon_\mathrm{F}=0  $};
 	\draw (1.4,0.38) node [font=\large] {c) $~\varepsilon_\mathrm{F}=0.3\,\varepsilon_0  $};
 	\draw (-1.15,-2.0) node [font=\large] { d) $~\varepsilon_\mathrm{F}=0.8\,\varepsilon_0 $};
 	\draw (1.4,-2.0) node [font=\large] { e) $~\varepsilon_\mathrm{F}=1.5\,\varepsilon_0 $};
 	\draw (-2.5,-0.55) node [font=\large] [rotate=90] {$ k_y/k_0 $};
 	\draw (-2.5,-3.1) node [font=\large] [rotate=90] {$ k_y/k_0 $};
 	\draw (-0.9,-4.57) node [font=\large] {$ k_x/k_0 $};
 	\draw (1.65,-4.57) node [font=\large] {$ k_x/k_0 $};
 	\end{tikzpicture}
 	\caption{a) Low-energy spectrum of a 2D tilted NLSM [in the units of $ \varepsilon_0=k_0^2/(2m) $], versus $ k_x/k_0$ for $k_y=0$. b-e) Fermi surfaces for b) $ \varepsilon_{\rm F}=0$, c) $ \varepsilon_{\rm F}=0.3\,\varepsilon_0$, d)  $ \varepsilon_\mathrm{F}=0.8\, \varepsilon_0$, and e)  $ \varepsilon_\mathrm{F}=1.5\,\varepsilon_0$. Electron and hole portions of the Fermi surfaces are identified via blue and red colors, respectively. 
	We have fixed the tilt velocity at $ u=0.3\, v_0$, with $v_0=k_0/m$ in all plots. \label{fig:dispersion}}
 \end{figure}

In Fig. \ref{fig:dispersion}, panel a) shows the energy dispersion of a 2D tilted NLSM, and panels b)-e) illustrate the Fermi surfaces for the intrinsic (i.e., un-doped) system and for different doping values.
A finite tilt partially lifts the valence band into the positive energies and pushes down portions of the conduction band into the negative energies. This leads to the simultaneous appearance of electron and hole pockets at the Fermi levels of the intrinsic as well as low-doped systems. 
We can readily find the density-of-states per unit area (DOS) from the energy dispersions,
$\rho({\varepsilon})=\rho_0\left[2-\Theta(|\varepsilon|-\varepsilon_m)\right]$, 
where $\rho_0=g m/(2\pi)$ is the DOS of an ordinary 2D electron gas~\cite{giuliani2005quantum}, with $g$ the total degeneracy factor, $ \Theta(x) $ is the Heaviside step function, and $\varepsilon_m=\varepsilon_0(1+ u^2/v_0^2)$ with $v_0=k_0/m$  is the energy corresponding to the extremum point of the conduction band, and $\varepsilon_0=k_0^2/(2m)$.
Note that the Fermi surface becomes a full disk for $|\varepsilon_{\rm F}|>\varepsilon_m$, where $\varepsilon_{\rm F}$ is the Fermi energy
(see, Fig. \ref{fig:dispersion}, panel e).
Considering only electron-doped (i.e., $ \varepsilon_{\rm F}>0 $) systems, just for the sake of definiteness, we can easily find the Fermi energy in terms of the extra carrier density 
\be
\varepsilon_{\rm F}=
\left\{
\begin{matrix}
n_e/(2\rho_0) & n_e\leq n_m\\
n_e/\rho_0-\varepsilon_m & n_e>n_m
\end{matrix}\right.,
\ee
where $n_e$ is the doped carrier concentration (measured with respect to the intrinsic limit, i.e. $\varepsilon_{\rm F}=0$), and $n_m=2\rho_0 \varepsilon_m$ is the density at $\varepsilon_{\rm F}=\varepsilon_m$.

\section{Non-interacting density-density response function}\label{sec:Polarizability}
When an electronic system is subjected to an external electromagnetic perturbation, the charge distribution changes and the system becomes polarized. The non-interacting density-density response function, or the polarizability of a 2D NLSM is given by the bare bubble diagram~\cite{giuliani2005quantum}
\begin{equation}
\Pi(\qv,\omega)=\dfrac{g}{S}\sum_{\kv s s^\prime}F_{s s^\prime}(\kv,\kv')\dfrac{f(\varepsilon_{\kv, s})-f(\varepsilon_{\kv', s^\prime})}{\hbar \omega+\varepsilon_{\kv, s}-\varepsilon_{\kv' ,s^\prime}+i0^+} \label{eq:polarizability},
\end{equation}
where $S$ is the sample area, $\kv'=\kv+\qv$, $ f(\varepsilon)=1/[\exp(\beta (\varepsilon-\mu)+1]$ is the Fermi-Dirac distribution function, with $ \beta=1/(k_{\rm B}T) $ the inverse temperature, and $ \mu $ the chemical potential, $ F_{s s^\prime}(\kv,\kv')=|\langle \kv,s|\kv',s'\rangle|^2 = \left[1+ss^\prime \sgn( k^2-1)\sgn( k'^2-1)\right]^2/4  $ is the form factor obtained from the overlap between eigenstates. Note that this form factor is either zero or one.

In the following subsections, we present analytic results for the real and imaginary parts of $\Pi(\qv,\omega)$  at different doping regimes, and at zero temperature, where $\left.f(\varepsilon)\right|_{T\to 0}=\Theta(\varepsilon_{\rm F}-\varepsilon)$.

\subsection{Polarizability of intrinsic 2D NLSM}\label{sec:Un-doped}
In an intrinsic NLSM the Fermi energy is zero, and the imaginary part of $\Pi(\qv,\omega)$ reads
\be\label{eq:Im-int}
\Im m\,\Pi (q,\omega)=\frac{2\rho_0k_0}{{ q}}\sum_{s=\pm 1} s\sqrt{\kappa^2-\nu_{s}^2}\Theta(\kappa^2-\nu_s^2),
\ee
where $\kappa=\sqrt{ \varepsilon_m/\varepsilon_0}=\sqrt{1+u^2/v_0^2}$, and $\nu_{\pm}={ \omega}/ (v_0 q)\pm  q/(2 k_0)$. 

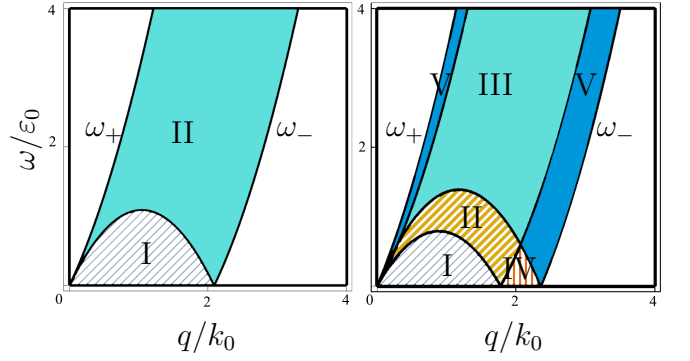
\begin{figure}
	\centering
	 \begin{tikzpicture}[scale=1.6]
	\node at (0,0) {\includegraphics[width=\linewidth]{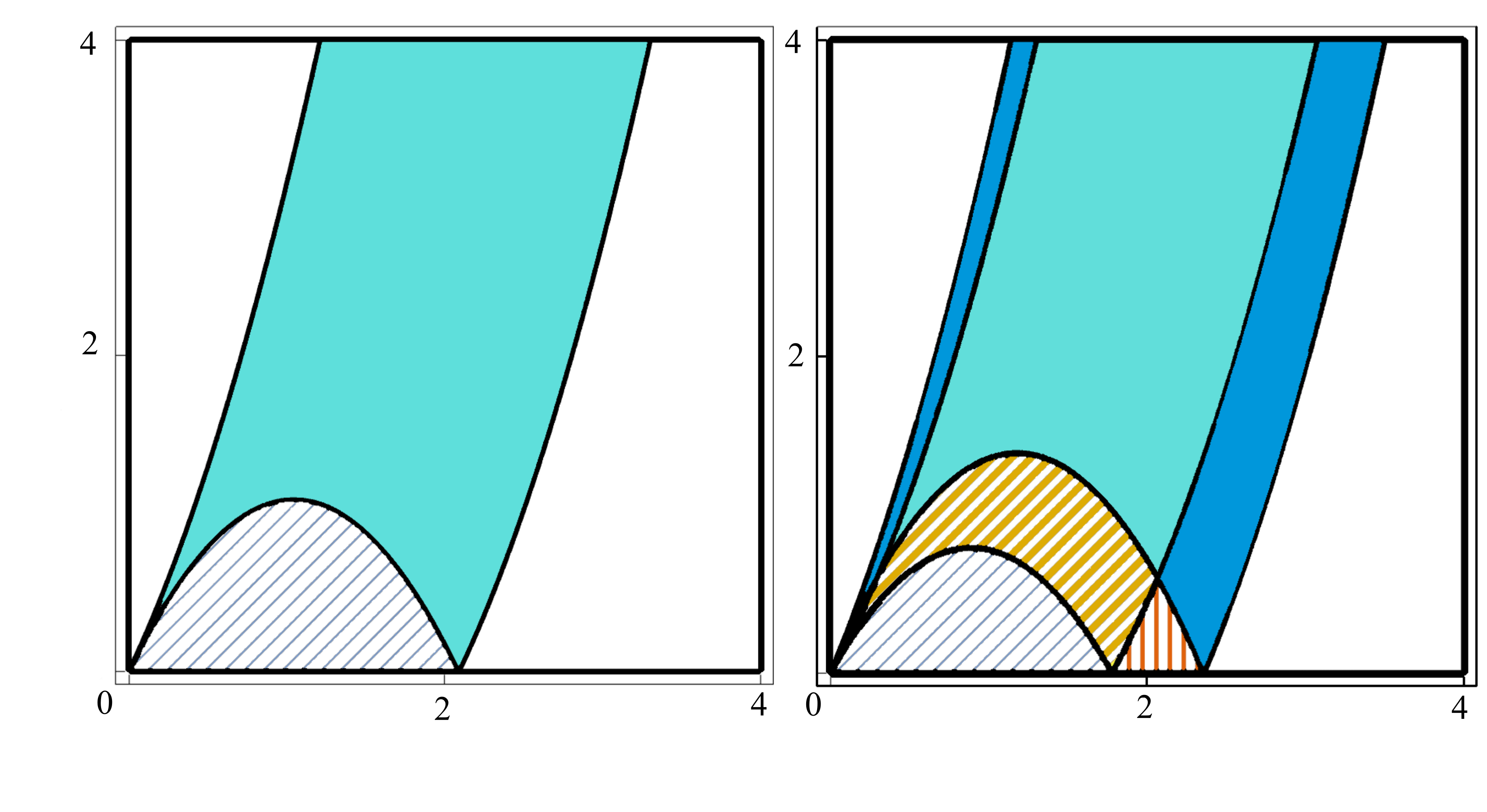}};
	\draw (-1.1,-1.4) node [font=\large] {$ q/k_0 $};
	\draw (1.45,-1.4) node [font=\large] {$ q/k_0 $};
	\draw (-2.6,0.25) node [font=\large] [rotate=90] {$ \omega/\varepsilon_0 $};
	\draw (-1.95,0.3) node [font=\large] {$ \omega_{+} $};
	\draw (-0.36,0.3) node [font=\large] {$ \omega_{-} $};
	\draw (0.55,0.3) node [font=\large] {$ \omega_{+} $};
	\draw (2.3,0.3) node [font=\large] {$ \omega_{-} $};
	\draw (-1.6,-0.7) node [font=\large] {$ \rm I $};
	\draw (-1.3,0.3) node [font=\large] {$ \rm II $};
	\draw (0.9,-0.8) node [font=\large] {$ \rm I $};
	\draw (1.1,-0.4) node [font=\large] {$\rm II $};
	\draw (1.3,0.7) node [font=\large] {$ \rm III $};
	\draw (1.5,-0.83) node [font=\large] {$ \rm IV $};
	\draw (2.05,0.7) node [font=\large] {$\rm V $};
	\draw (0.85,0.7) node [font=\large] {$ \rm V$};
	\end{tikzpicture} 
	\caption{ Electron-hole continuum of a 2D tilted NLSM with $ u=0.3\, v_0$, in the intrinsic regime (left) and low-doped regime with $\varepsilon_{\rm F}=0.3 \, \varepsilon_0$ (right). Different colors point to regions of the $\omega-q$ plane with different expressions for the imaginary part of the non-interacting polarizability (see the text for definitions), while in the white regions $ \Im m\,\Pi(q,\omega) $ vanishes.  \label{fig:continuum}}
\end{figure}

In the left panel of Fig. \ref{fig:continuum}, the shaded regions show the electron-hole continuum (EHC), where $ \Im m\,{\Pi}(q,\omega)$ is different from zero. $ \omega_{\pm}=q^2/(2 m) \pm \kappa v_0 q$ separates EHC from $ \Im m\,{\Pi}(q,\omega)= 0 $ regions, where single electron-hole excitation is not allowed.

Regions of the EHC labeled as I and II in the left panel of Fig. \ref{fig:continuum} refer to areas of the $\omega-q$ plane defined through
\be
\left\{
\begin{matrix}
{\rm I}:  &  \nu_{+}^2<\kappa^2 ;  ~\nu_{-}^2<\kappa^2\\
{\rm II}: & \nu_{+}^2>\kappa^2 ;  ~\nu_{-}^2<\kappa^2
\end{matrix}\right. .
\ee
With the help of the Kramers-Kronig relation, we can find the real part of the polarizability 
\be
\begin{split}
\Re e\,\Pi&(q,\omega)=  -2\rho_0\\
&+\frac{2\rho_0 k_0}{{ q}}\sum_{s=\pm 1} s\sgn(\nu_{s})\sqrt{\nu_{s}^2-\kappa^2}\Theta(\nu_{s}^2-\kappa^2).
\end{split}
\label{eq:Re-int}
\ee

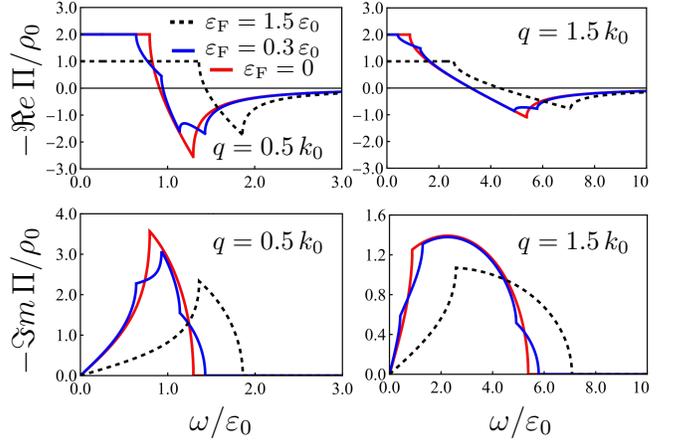
\begin{figure}
	\centering
	\begin{tikzpicture}[scale=0.7]
	\node at (0,0) {\includegraphics[width=0.98\linewidth]{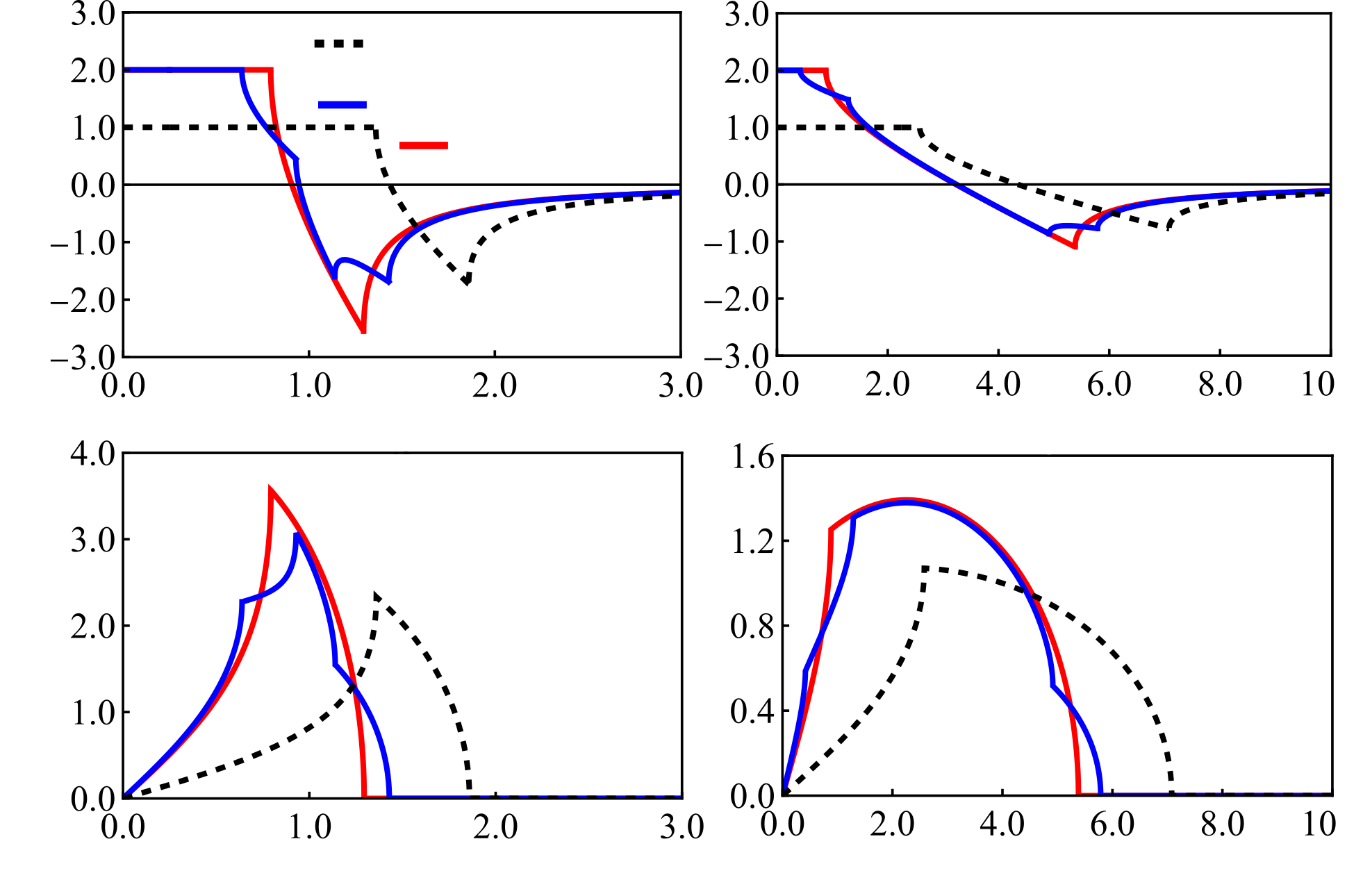}};
	\draw (-2.3,-4) node [font=\large] {$ \omega/\varepsilon_0 $};
	\draw (3.4,-4) node [font=\large] {$  \omega/\varepsilon_0 $};
	\draw (-6,2.3) node [font=\large] [rotate=90] {$- \Re e\, {\Pi} /\rho_0 $};
	\draw (-6,-1.7) node [font=\large] [rotate=90] {$ -\Im m\, {\Pi} /\rho_0 $};
	\draw (-1.2,2.7) node  [font=\small] {$\varepsilon_{\rm F}=0 $};
	\draw (-1.5,3.1) node  [font=\small] {$\varepsilon_{\rm F}=0.3\, \varepsilon_0 $};
	\draw (-1.5,3.6) node  [font=\small] {$\varepsilon_{\rm F}=1.5\, \varepsilon_0 $};
	\draw (-1.4,1.2) node [font=\normalsize] {$q=0.5\,k_0 $};
	\draw (4.4,3.4) node [font=\normalsize] {$q=1.5\,k_0 $};
	\draw (-1.4,-0.6) node [font=\normalsize] {$q=0.5\,k_0 $};
	\draw (4.4,-0.6) node [font=\normalsize] {$q=1.5\,k_0 $};
	\end{tikzpicture} 
	\caption{The real (top panels) and imaginary (bottom panels) parts of the non-interacting dynamical polarizability of a 2D tilted NLSM as a function of $\omega/\varepsilon_0$, for $ q=0.5 \,k_0 $ (left panels) and $ q=1.5\, k_0 $ (right panels) and for varying Fermi energies. The Fermi energies $\varepsilon_{\rm F}=0$, $\varepsilon_{\rm F}=0.3\, \varepsilon_0$, and $\varepsilon_{\rm F}=2.0\, \varepsilon_0 $, correspond to the intrinsic, low-doped, and high-doped regimes, respectively. The tilt velocity is fixed at $u=0.3\,v_0 $ in all panels.}
	\label{fig:Re-Im}
\end{figure}  

\subsection{Polarizability of 2D NLSM at low dopings}\label{sec:Low-doping}
For $0<\varepsilon_{\rm F}<\varepsilon_m$, the Fermi surface has a hollow in its center. In this regime, analytic expressions for the imaginary and real parts of the zero temperature density-density linear response function are obtained respectively as
\be\label{eq:Im-low}
\Im m \,{\Pi} (q,\omega)=\dfrac{\rho_0k_0}{{ q}}\sum_{s,s'=\pm 1}
s\sqrt{\kappa^2_{s'}-\nu_{s}^2}\Theta(\kappa^2_{s'}-\nu_{s}^2),
\ee
and
\be\label{eq:Re-low}
\begin{split}
\Re e\,&{\Pi}(q,\omega)=-2\rho_0\\
&+\dfrac{\rho_0 k_0}{{ q}}
\sum_{s,s'=\pm 1} s\sgn(\nu_s)\sqrt{\nu_{s}^2-\kappa^2_{s'}}\Theta(\nu_{s}^2-\kappa^2_{s'}),
\end{split}
\ee
where $\kappa_\pm=\sqrt{\kappa^2\pm  \varepsilon_{\rm F}/\varepsilon_0}$. 

The right panel of Fig. \ref{fig:continuum}, illustrates the EHC of a 2D tilted NLSM in the low-doped regime.
Now, the boundaries of the EHC are given by $\omega_{\pm}=q^2/(2 m) \pm \kappa_\pm v_0 q$.
Different regions of the EHC are defined through the following conditions
\be
\left\{
\begin{matrix}
{\rm I}: &  \nu_{+}^2<\kappa^2_- ;~ \nu_{-}^2<\kappa^2_+;~ \nu_{-}^2<\kappa^2_-; \nu_{+}^2<\kappa^2_+  \\
{\rm II}: &  \nu_{+}^2>\kappa^2_- ;~ \nu_{-}^2<\kappa^2_+;~ \nu_{-}^2<\kappa^2_-  ;~ \nu_{+}^2<\kappa^2_+ \\
{\rm III}: &  \nu_{+}^2>\kappa^2_-  ;~ \nu_{-}^2<\kappa^2_+  ~ \nu_{-}^2<\kappa^2_-  ;~ \nu_{+}^2>\kappa^2_+ \\
{\rm IV}: & \nu_{+}^2>\kappa^2_-  ;~ \nu_{-}^2>\kappa^2_+  ;~ \nu_{-}^2<\kappa^2_-  ;~ \nu_{+}^2<\kappa^2_+ \\
{\rm V}: & \nu_{+}^2>\kappa^2_-  ;~ \nu_{-}^2>\kappa^2_+  ;~ \nu_{-}^2<\kappa^2_-  ;~ \nu_{+}^2>\kappa^2_+ 
\end{matrix}
\right. .
\ee

\subsection{Polarizability of 2D NLSM at high dopings}\label{sec:high-doping}
In the high doping regime, where $ \varepsilon_\mathrm{F}>\varepsilon_m$, the Fermi surface becomes a filled circle with radius $k_0 \kappa_+$, centered around $(-m u,0)$ in the $k_x-k_y$ plane.
The expression for the density-density response in this regime is similar to the intrinsic case, and one only needs to substitute $\kappa$ in Eqs.~\eqref{eq:Im-int} and \eqref{eq:Re-int} with $\kappa_+$ and divide the whole expressions with 2, as the density of states in this case is half of the un-doped system. We write the final expressions here just for the sake of completeness
\be\label{eq:Im-high}
\Im m\,\Pi (q,\omega)=\frac{\rho_0k_0}{{ q}}\sum_{s=\pm 1} s\sqrt{\kappa^2_+-\nu_{s}^2}\Theta(\kappa^2_+-\nu_s^2).
\ee
\be
\begin{split}
\Re e\,&\Pi(q,\omega)=  -\rho_0\\
&+\frac{\rho_0k_0}{{ q}}\sum_{s=\pm 1} s\sgn(\nu_{s})\sqrt{\nu_{s}^2-\kappa_+^2}\Theta(\nu_{s}^2-\kappa_+^2).
\end{split}
\label{eq:Re-high}
\ee
It is clear from Eq.~\eqref{eq:Im-high} that the EHC looks similar to the left panel of Fig.~\ref{fig:continuum}, where the boundaries now depend on the Fermi energy
$ \omega_{\pm}=q^2/(2 m) \pm \kappa_+ v_0 q$, and different regions of the EHC are defined through
\be
\left\{
\begin{matrix}
{\rm I}:  &  \nu_{+}^2<\kappa^2_+ ;  ~\nu_{-}^2<\kappa^2_+\\
{\rm II}: & \nu_{+}^2>\kappa^2_+ ;  ~\nu_{-}^2<\kappa^2_+
\end{matrix}
\right. .
\ee 

Fig. \ref{fig:Re-Im} illustrates the behavior of non-interacting dynamical polarizability ${\Pi}(q,\omega)$ versus $\omega$, for varying wave vectors and doping regimes.
In the intrinsic limit, the real part of the polarizability (red curves in the top panels of Fig. \ref{fig:Re-Im}) has a negative plateau for $ \omega<\omega_{-}$, and changes sign at higher frequencies. The imaginary part of the polarizability (red curves in the bottom panels of Fig. \ref{fig:Re-Im}) for $q<2 k_0$ displays a cusp at the boundary between regions $ \rm I $ and $\rm II $ of the EHC, and vanishes for $ \omega>\omega_{+}$, i.e., beyond the upper edge of the EHC.
In the low-doped regime (blue curves in Fig. \ref{fig:Re-Im}), both the real and imaginary parts of the polarizability display several distinct cusps corresponding to crossings between different regions within the EHC. The imaginary part vanishes outside the EHC, i.e., for $ \omega>\omega_{+}$.
The behavior of $\Pi(\qv,\omega)$ in the high-doped regime (dashed black lines), apart from the Fermi energy dependence, is similar to the intrinsic system.

\section{Static response and Friedel oscillations}\label{sec:Static screening}
In this section, we discuss the behavior of the non-interacting linear density-density response function of a 2D tilted NLSM in the static limit and investigate how this system screens charged impurities at different regimes of doping. 
\subsection{Static polarizability}\label{sec:Static}
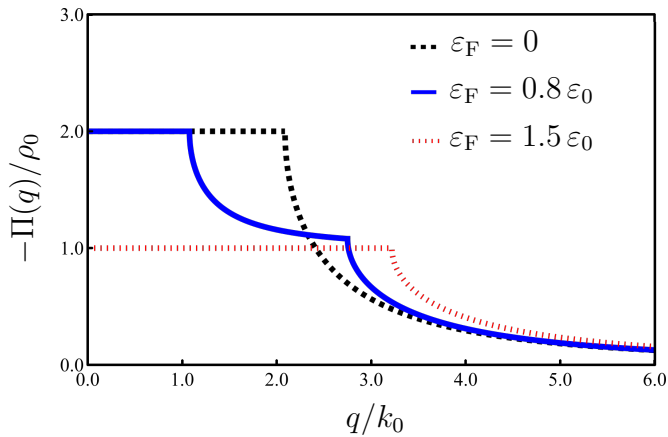
\begin{figure}
	\centering
		\begin{tikzpicture}[scale=1.2]
	\node at (0,0) {\includegraphics[width=0.97\linewidth]{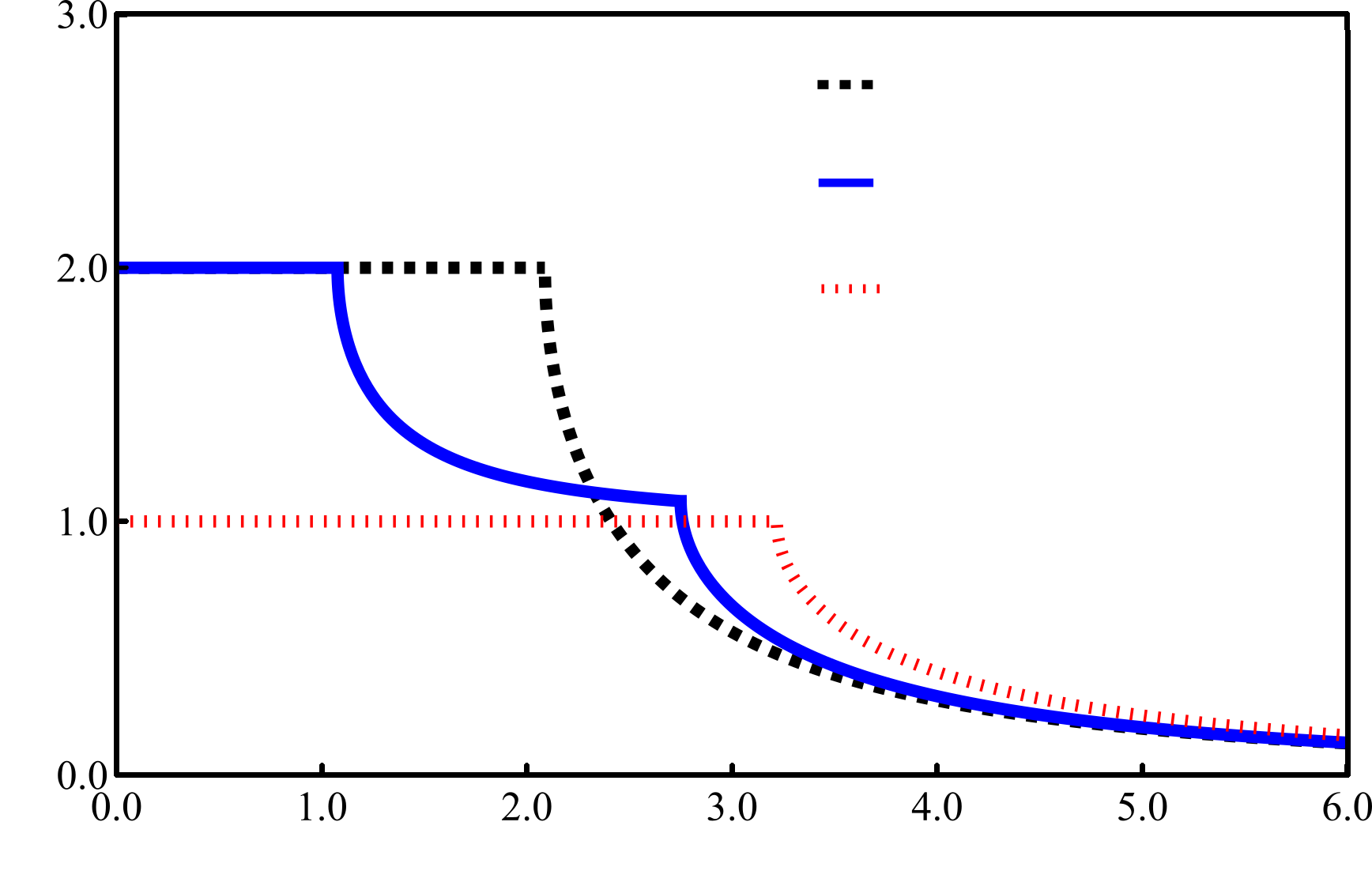}};
	\draw (0.28,-2.3) node [font=\large] {$ q/k_0 $};
	\draw (-3.6,0.15) node [font=\large] [rotate=90] {$ -\Pi(q)/\rho_0 $};
	\draw (1.6,1.87) node  {\large $ \varepsilon_{\mathrm{F} }=0 $};
	\draw (1.9,1.35) node  {\large$  \varepsilon_{\mathrm{F} }=0.8\,\varepsilon_{0} $};
	\draw (1.9,0.82) node  {\large $ \varepsilon_{\mathrm{F} }=1.5\, \varepsilon_{0} $};
	\end{tikzpicture}  
	\caption{Behavior of the non-interacting static polarizability of a 2D tilted NLSM (in the units of $-\rho_0$) versus $ q/k_0 $, in the intrinsic (dashed black line), low-doped with $\varepsilon_\mathrm{F}=0.8\, \varepsilon_{0} $ (solid blue line), and high-doped with $\varepsilon_\mathrm{F}=1.5 \, \varepsilon_{0}  $ (dotted red line) regimes. The tilt parameter is fixed at $ u=0.3\, v_0$.
		\label{fig:static}}
\end{figure}

The static polarizability is a real function, and its analytic form at different dopings is obtained easily from the $\omega\to 0$ limit of Eqs.~\eqref{eq:Re-int}, \eqref{eq:Re-low}, and \eqref{eq:Re-high}, that reads
\begin{widetext}
\be\label{eq:static}
\Pi(q)=-\rho_0\left\{
\begin{matrix}
2-2\Theta(q-2k_0 \kappa ) \sqrt{1-(2 k_0\kappa/q)^2}, & \varepsilon_{\rm F}=0\\
2-\sum_{s=\pm 1} \Theta(q-2 k_0\kappa_{s})\sqrt{1-(2 k_0\kappa_s /q)^2}, & 0<|\varepsilon_{\rm F}|<\varepsilon_m\\
1-\Theta(q-2k_0 \kappa_+ ) \sqrt{1-(2k_0\kappa_+ /q)^2}, & |\varepsilon_{\rm F}|>\varepsilon_m
 \end{matrix}
 \right. .
 \ee
 \end{widetext}
In the long wavelength limit, the static density-density response function is equal to the negative of the density of states at the Fermi level.  
Another important characteristic of static polarizability is its singular behavior due to the Fermi surface that separates empty states from filled ones. 
As it is shown in Fig. \ref{fig:static}, static polarizability of the intrinsic and highly doped systems have singular points at $q= k_0\kappa $, and $q=k_0\kappa_+ $, respectively. In the low-doping regime, on the other hand, two singularities at $q=k_0 \kappa_\pm$ correspond to the radius of inner and outer circles of the Fermi surfaces (see, panels c and d in Fig.~\ref{fig:dispersion} ). 

\subsection{Friedel oscillations}\label{sec:Friedel oscillations}
The singular points of the static density response function cause Friedel oscillations in the impurity screening. 
The Fourier transform of the static polarizability, i.e., 
$\Pi(r)=\int {\rm d}^2q\, \Pi(\qv)e^{\mathrm{i}\qv.\rv} /(2\pi)^2$,
gives the charge density induced at a distance $r$ from a point impurity potential~\cite{giuliani2005quantum}.
Taking the Fourier transforms of Eq.~\eqref{eq:static}, we find
\begin{widetext}
\be\label{eq:chi_r}
\Pi(r)=\frac{8k_0^2\rho_0}{\pi} \sum_{n=0,1}
\left\{
\begin{matrix}
 2\kappa^2 J_{n}(2k_0\kappa  r)N_{n}(2k_0\kappa  r), & \varepsilon_{\rm F}=0\\
\sum_{s=\pm 1} \kappa_s^2 J_{n}(2k_0\kappa_s  r)N_{n}(2k_0\kappa_s  r)  , & 0<|\varepsilon_{\rm F}|<\varepsilon_m\\
\kappa_+^2 J_{n}(2k_0\kappa_+  r)N_{n}(2k_0\kappa_+  r), & |\varepsilon_{\rm F}|>\varepsilon_m
\end{matrix}
\right. ,
\ee
\end{widetext}
where $J_{n}(x)$ and $N_{n}(x)$ are the Bessel functions of the first and second kind, respectively. 

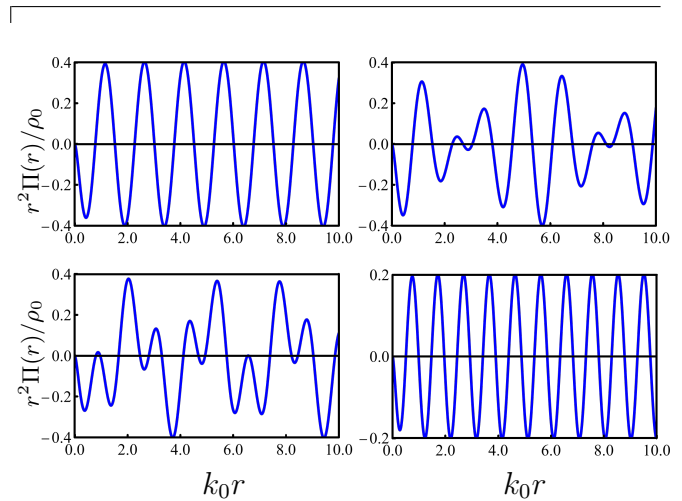
\begin{figure}
	\centering
	\begin{tikzpicture}[scale=1.6]
	\node at (0,0) {\includegraphics[width=\linewidth]{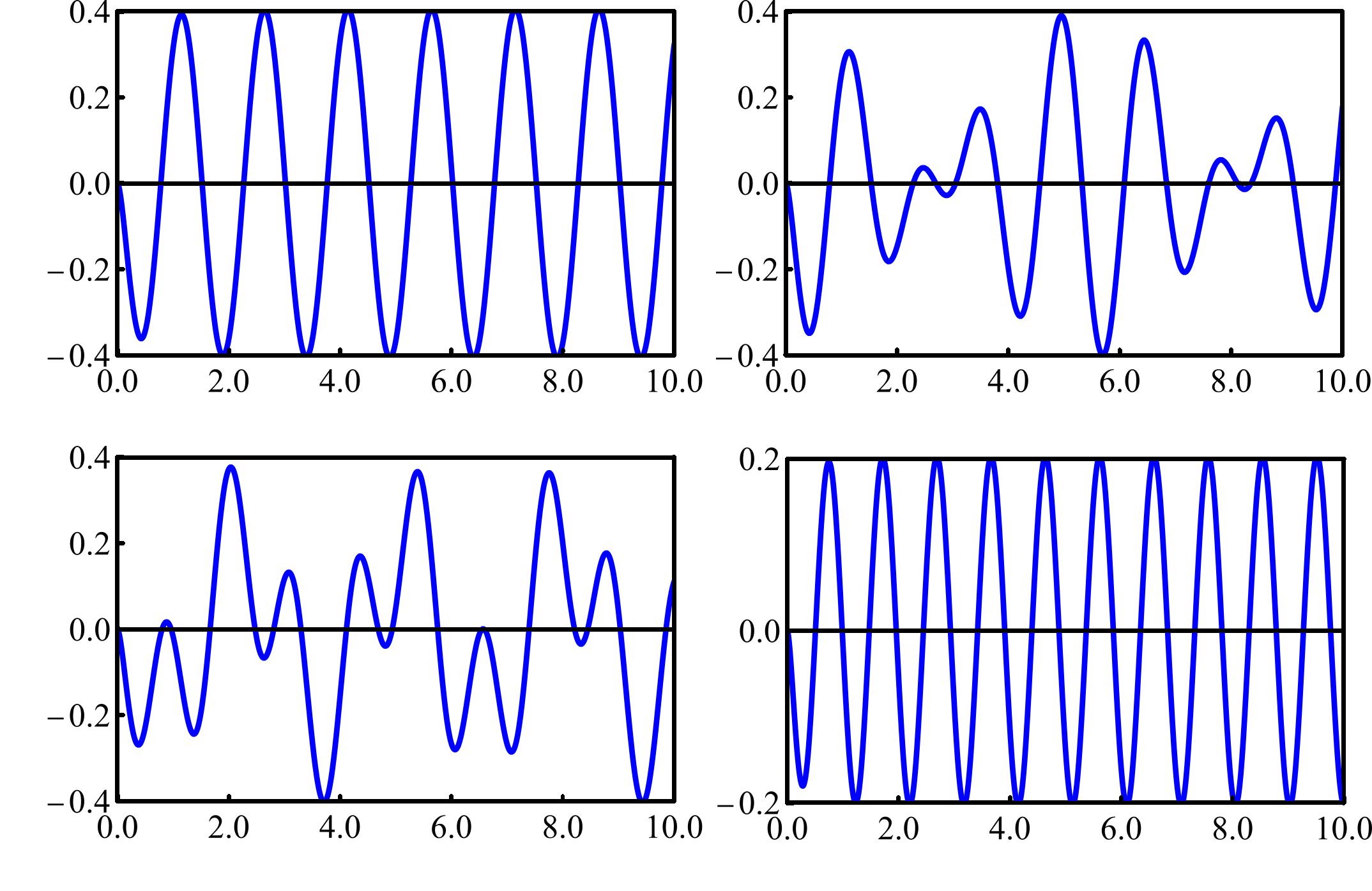}};
	\draw (-1,-1.8) node [font=\large] {$ k_0r $};
	\draw (1.5,-1.8) node [font=\large] {$ k_0r $};
	\draw (-2.6,0.9) node [rotate=90] {$ r^2\Pi(r)/\rho_0 $};
	\draw (-2.6,-0.7) node [rotate=90] {$ r^2\Pi(r)/\rho_0 $};
	\end{tikzpicture}  
	\caption{The real space static polarizability $\Pi(r)$ (in the units of $\rho_0/r^2$) of a 2D tilted NLSM versus $k_0 r$, at  $\varepsilon_\mathrm{F}=0 $ (top left), $\varepsilon_\mathrm{F}= 0.3\,  \varepsilon_{0}$ (top right), $\varepsilon_\mathrm{F}= 0.8\,  \varepsilon_{0}$ (bottom left), $\varepsilon_\mathrm{F}= 1.5\,  \varepsilon_{0}$ (bottom right). The tilt parameter is fixed at $ u=0.3\, v_0$ in all figures.
		\label{fig:Friedel}}
\end{figure}

It is easy to verify that $\Pi(r)$ decays as $ r^{-2} $ at long distances. Furthermore, the Bessel functions give rise to an oscillatory spatial modulation of the induced density. 
The behavior of $\Pi(r)$ versus $r$ at different Fermi energies are depicted in Fig.~\ref{fig:Friedel}. The period of oscillations in the intrinsic regime is given by the radius of the nodal ring $k_0$, while in the doped systems, both the ring radius and the Fermi energy contribute to the period. At low dopings, two distinct singularities of $\Pi(q)$ give rise to the superposition of two oscillating functions with different periods. The resulting beat patterns are evident in the top-right and bottom-left panels of Fig.~\ref{fig:Friedel}.

\section{Drude weight and plasmon dispersion}\label{sec:Plasmons}
In the random phase approximation, the interacting linear density-density response function can be expressed as 
\begin{equation}
\chi^{\rm RPA}(q,\omega)=  \dfrac{\Pi(q,\omega)}{{\varepsilon}^{\rm RPA}(q,\omega)}\label{eq:rpa}.
\end{equation}
Here, ${\varepsilon}^{\rm RPA}(q,\omega)=1-v(q){\Pi}(q,\omega)$ is the dynamical dielectric function within the RPA, where
$v(q)=2\pi e^2/q$ is the Fourier transform of the Coulomb interaction in 2D, with $e$ the charge of electron. 
Zeros of the dynamical dielectric function (or equivalently, poles of the interacting dynamical polarizability) give the dispersions of the collective density oscillations, i.e., plasmon modes. 

To find the dispersion of undamped plasmons, we solve $1-v(q)\Re e\, \Pi(q,\omega)=0$, outside the EHC.
 In the following, we first investigate the long-wavelength behavior of plasmon modes in a 2D tilted NLSM at different doping regimes. 
In the $q\to 0$ limit, from~\cite{abedinpour2011drude}
\be
\lim_{\omega \to 0}\lim_{q \to 0}\Re e\,\Pi(q,\omega)=  \dfrac{\cal D}{\pi e^2}\dfrac{q^2}{\omega^2}\label{eq:plasmon-Drude},
\ee
we find
\be
\omega_{\rm pl}(q\to 0)\approx\sqrt{2 {\cal D} q},
\ee
where 
\be
{\cal D}=
{\cal D}_0
\left\{
\begin{matrix}
2\kappa^2, & |\varepsilon_{\rm F}|\leq\varepsilon_m\\
\kappa_+^2, & |\varepsilon_{\rm F}|\geq\varepsilon_m
\end{matrix}
\right. ,
\label{eq:Drude}
\ee
is the Drude weight with  $ {\cal D}_0=g e^2  \varepsilon_0 /2$. 

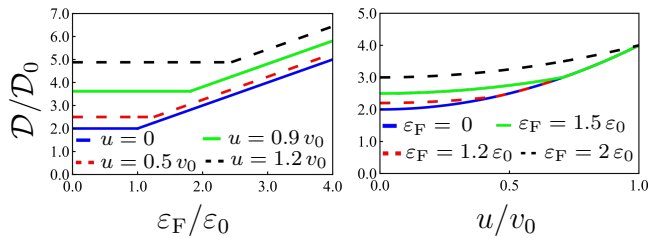
\begin{figure}
	\centering
	\begin{tikzpicture}[scale=1.6]
	\node at (0,0) {\includegraphics[width=0.95\linewidth]{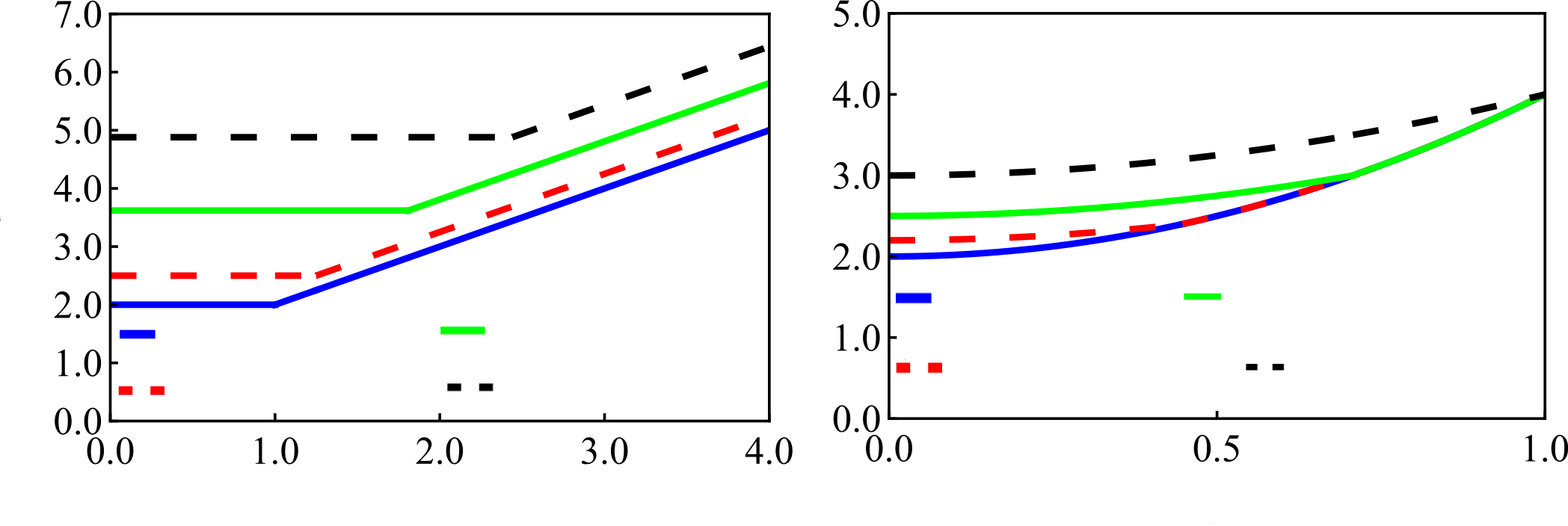}};
	\draw (-1.75,-0.22)[font=\footnotesize] node {$ u=0 $};
	\draw (-1.57,-0.42)[font=\footnotesize]  node  {$ u=0.5\, v_0 $};
	\draw (-0.51,-0.22)[font=\footnotesize] node {$ u=0.9\, v_0  $};
	\draw (-0.49,-0.4)[font=\footnotesize]  node  {$ u=1.2\, v_0 $};
	\draw (-2.6,0.1) node [rotate=90] {\large $ {\cal D}/{\cal D}_0$};
	\draw (-1.2,-0.9) node [font=\large] {$ \varepsilon_\mathrm{F}/\varepsilon_{0} $};
	\draw (1.4,-0.9) node [font=\large] {$ u/v_0 $};
	\draw (0.83,-0.12)[font=\footnotesize] node  {$  \varepsilon_{\mathrm{F} }= \,0 $};
	\draw (1.01,-0.34)[font=\footnotesize] node  {$ \varepsilon_{\mathrm{F} }=1.2\, \varepsilon_{0} $};
	\draw (1.95,-0.1)[font=\footnotesize] node  {$ \varepsilon_{\mathrm{F} }=1.5\, \varepsilon_{0} $};		
	\draw (2.09,-0.34)[font=\footnotesize] node  {$ \varepsilon_{\mathrm{F} }=2\, \varepsilon_{0} $};
	\end{tikzpicture}  
	\caption{The Drude weight (in the units of $ {\cal D}_0=g e^2 \varepsilon_0/2 $) versus Fermi energy, for varying values of the tilt velocity (left) and versus tilt velocity for varying Fermi energies (right).
\label{fig:Drude}}
\end{figure}

Fig. \ref{fig:Drude} presents the Fermi energy and tilt velocity dependence of the Drude weight.  
As it is evident from the left panel of  Fig. \ref{fig:Drude}, the Drude weight is constant for $|\varepsilon_{\rm F}|<\varepsilon_m  $ and linearly increases with Fermi energy for  $|\varepsilon_{\rm F}|>\varepsilon_m$.
The right panel of Fig. \ref{fig:Drude} shows how the Drude weight enhances with increasing the tilt velocity at different Fermi energies. 
Notice that the low doping regime is defined as $|\varepsilon_{\rm F}|<\varepsilon_m=\varepsilon_0(1+u^2/v_0^2)$. Increasing the tilt velocity, more curves fall into this low doping regime and merge with the blue line that shows the intrinsic system behavior. 

The Drude weight also manifests itself in the behavior of optical conductivity in the local (i.e., $q\to 0$) limit. 
The only contribution to the optical conductivity of our model Hamiltonian~\eqref{eq:hamil} arises from the intra-band transitions~\cite{barati2017optical}.
Therefore the real part of the optical conductivity of a clean system only has the Drude peak, i.e., 
$\Re e\, \sigma(\omega)={\cal D}\delta(\omega)$, and its imaginary part reads $\Im m\, \sigma(\omega)={\cal D}/(\pi\omega)$.
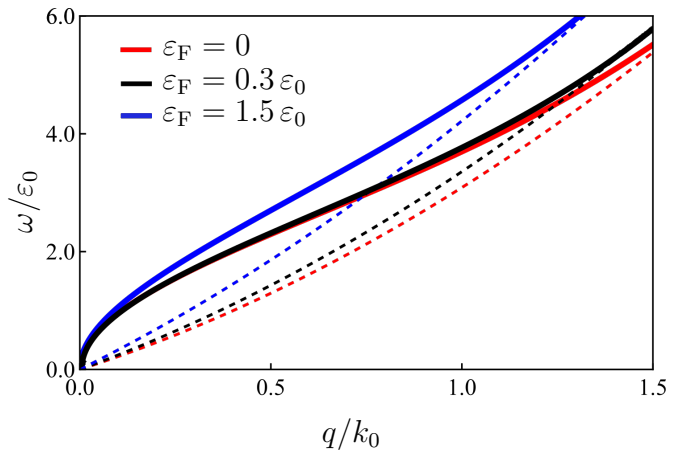
\begin{figure}
	\centering
	\begin{tikzpicture}[scale=0.5]
	\node at (0,0) {\includegraphics[width=\linewidth]{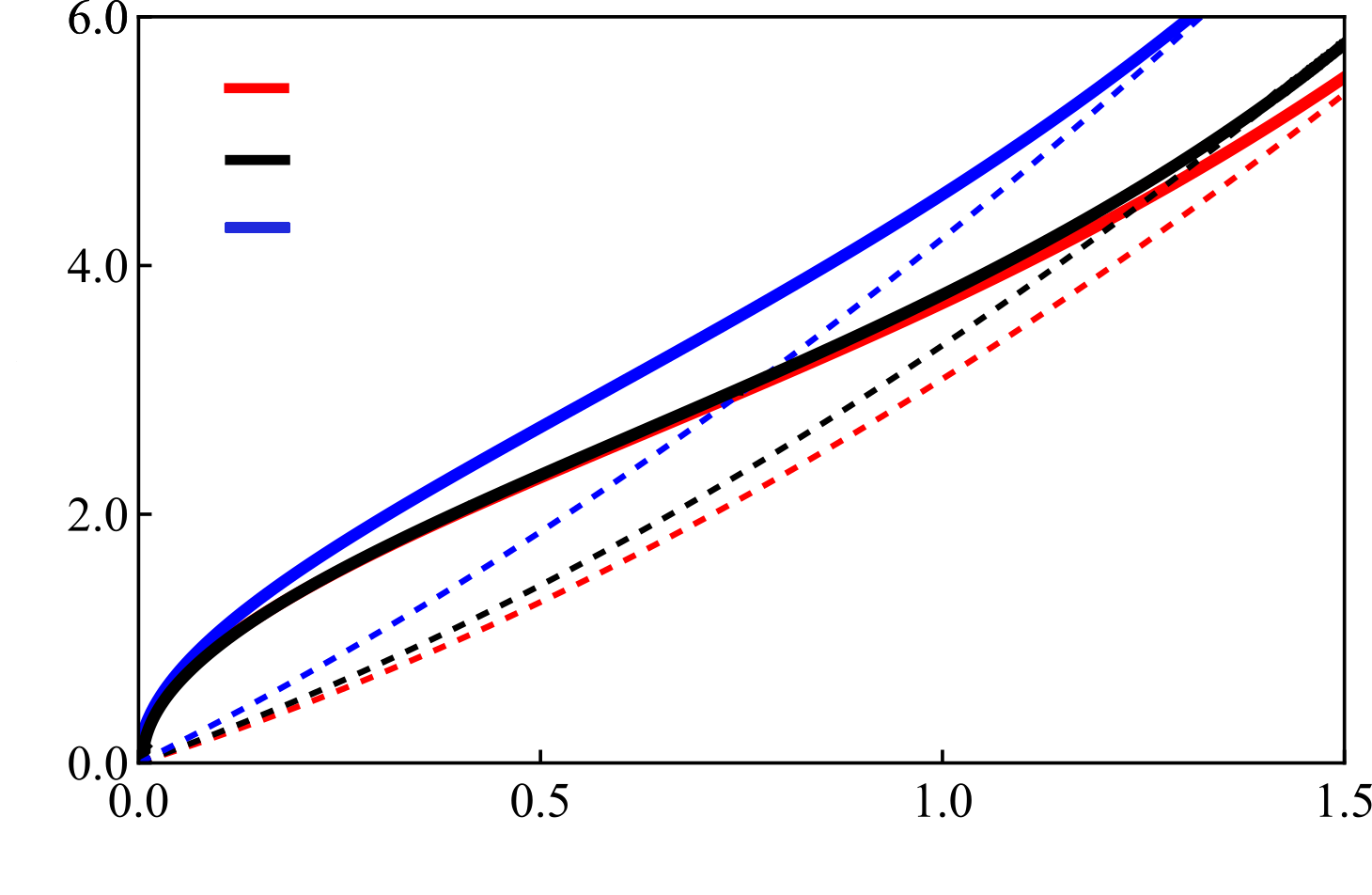}};
	\draw (0.3,-5.7) node [font=\large] {$ q/k_0 $};
	\draw (-8.4,0.5) node [font=\large] [rotate=90] {$ \omega/\varepsilon_0 $};
	\draw (-3.55,4.6) node  {\large $ \varepsilon_{\mathrm{F} }=0 $};
	\draw (-2.8,3.7) node  {\large $  \varepsilon_{\mathrm{F} }=0.3\,\varepsilon_{0} $};
	\draw (-2.8,2.85) node  {\large $ \varepsilon_{\mathrm{F} }=1.5\, \varepsilon_{0} $};
	\end{tikzpicture} 
	\caption{Plasmon dispersion (in the units of $\varepsilon_0$) versus $q/k_0$ for intrinsic $\varepsilon_\mathrm{F}= 0  $ (solid red line), low-doping with $\varepsilon_\mathrm{F}= 0.3\, \varepsilon_{0}  $ (solid black line), and high-doping with $\varepsilon_\mathrm{F}=1.5\, \varepsilon_{0}  $ (solid blue line) 2D tilted NLSMs. The tilt velocity is fixed at $u=0.3\, v_0$ and  $\eta=2g/(k_0 a_{\rm B})=4$.
 The upper edges of the EHC at each Fermi energy are shown with dashed lines of the same color as their corresponding plasmon dispersion curves.
 \label{fig:plasmon}}.
\end{figure} 

As we noted, the Drude weight, and therefore the plasmon dispersion to the leading order in wave vector, is interestingly independent of the carrier concentration below a threshold density $n_m=2\rho_0 \varepsilon_{m}$. 
However, the effects of Fermi energy on the plasmon dispersion show up at larger wave vectors. 
Next to the leading order in $q$, the plasmon dispersion reads
\be\label{eq:wpl_q2}
\omega_{\rm pl}(q\to 0) \approx \sqrt{2 {\cal D}q}\left(1+\gamma q+\cdots \right),
\ee
with
\be
\gamma=\frac{3 a_{\rm B}}{8 g} 
\left\{
\begin{matrix}
1+\left(\varepsilon_{\rm F }/\varepsilon_m\right)^2~, & |\varepsilon_{\rm F}|\leq \varepsilon_m\\
2~, & |\varepsilon_{\rm F}|>\varepsilon_m
\end{matrix}
\right. ,
\ee
where $a_{\rm B}=1/(me^2)$ the effective Bohr radius. 
In passing, we note that two widely explored examples of two-dimensional electronic systems, i.e., the ordinary 2D electron gas with a single parabolic band  (2DEG), and the Dirac fermions in a single layer doped graphene sheet, also have a long wavelength plasmon dispersion similar to Eq.~\eqref{eq:wpl_q2}.
The non-interacting Drude weight in both of these systems is proportional to the Fermi energy, i.e., ${\cal D}=e^2 \varepsilon_{\rm F}$, the behavior we recover only at high dopings for 2D tilted NLSMs. 
The coefficient of the sub-leading correction to the plasmon dispersion of 2DEG and graphene are $\gamma_{\rm 2DEG}=3 a_{\rm B}/(4 g)$ and $\gamma_{\rm G}=-g e^2 \varepsilon_{\rm F}/v_{\rm F}^2$, respectively, where $v_{\rm F}$ is the energy independent Fermi velocity of graphene~\cite{hwang2007dielectric,stern1967polarizability}.
 
It is possible to find analytic expressions for the plasmon dispersion of 2D tilted NLSMs at arbitrary wave vectors, in the intrinsic and highly doped limits as
\be
\omega_{\rm pl}(q)=\varepsilon_{0} \dfrac{\sqrt{{\bar q}}( \eta+ {\bar q})\sqrt{4\eta^2\kappa^2+2 \eta {\bar q^3}+\bar q^4 }}{\eta \sqrt{2\eta +\bar q}}\label{eq:plasmon-int},
\ee
and
\be
\omega_{\rm pl}(q)=\varepsilon_{0} \dfrac{\sqrt{{\bar q}}( \eta+ {2\bar q})\sqrt{\eta^2\kappa_{+}^2+ \eta {\bar q^3}+\bar q^4 }}{\eta \sqrt{\eta +\bar q}}\label{eq:plasmon-high},
\ee
respectively, where ${\bar q}=q/k_0$ and $\eta=2g/(k_0 a_{\rm B})$. 
Fig.~\ref{fig:plasmon} displays the full plasmon dispersions in three different doping regimes within the RPA.
Note that for intermediate values of the Fermi energies, we were not able to find an analytic expression for the plasmon dispersion at arbitrary wave vectors. Therefore the plasmon dispersion at $\varepsilon_{\mathrm{F} }=0.3\,\varepsilon_{0}$ (solid black line in Fig.~\ref{fig:plasmon}) is obtained numerically.

\section{Summary and Conclusion}\label{sec:summary}
We have investigated the linear density-density response of two-dimensional tilted nodal-line semimetals in the intrinsic and doped regimes. Despite the anisotropic band structure of the system, integration over the angular part of the wave vector $\kv$ washes out the anisotropy of polarizability in our simple model. Correspondingly, screening, collective mode dispersion, and optical conductivity all remain isotropic at all doping levels.

At low dopings, two distinct singularities in the static polarizability give rise to beating patterns in the Friedel oscillations. 
The period of oscillation in the intrinsic limit is given by the nodal ring radius while in the doped systems, both the Fermi energy and the radius of the nodal ring contribute to the oscillation period. 
We also find analytical expressions for the plasmon dispersion in the intrinsic and highly-doped systems. The tilt strength enhances the plasmon frequency. At long wavelengths, the collective mode frequency is proportional to $q^{1/2} $. However, the Drude weight is independent of the carrier density below a threshold carrier density.

\bibliography{Tilted_nodal.bbl}

\end{document}